\documentclass[a4paper,11pt]{article}
\pdfoutput=1
\usepackage{jinstpub}
\usepackage{lineno}
\input{dsnew.def}
\newcommand{\Alberta}{Department of Physics, University of Alberta, Edmonton, AB T6G 2R3, Canada}
\newcommand{\APC}{APC, Universit\'e Paris Diderot, CNRS/IN2P3, CEA/Irfu, Obs de Paris, USPC, Paris 75205, France}
\newcommand{\AQLNGS}{INFN Laboratori Nazionali del Gran Sasso, Assergi (AQ) 67100, Italy}
\newcommand{\AQGSSI}{Gran Sasso Science Institute, L'Aquila 67100, Italy}
\newcommand{\AstroCeNT}{AstroCeNT, Nicolaus Copernicus Astronomical Center, Polish Academy of Sciences, Rektorska 4, 00-614 Warsaw, Poland}
\newcommand{\Augustana}{Physics Department, Augustana University, Sioux Falls, SD 57197, USA}
\newcommand{\Belgorod}{Radiation Physics Laboratory, Belgorod National Research University, Belgorod 308007, Russia}
\newcommand{\BHSU}{School of Natural Sciences, Black Hills State University, Spearfish, SD 57799, USA}
\newcommand{\BINP}{Budker Institute of Nuclear Physics, Novosibirsk 630090, Russia}
\newcommand{\BNLaddress}{Brookhaven National Laboratory, Upton, NY 11973, USA}
\newcommand{\BOINFN}{INFN Bologna, Bologna 40126, Italy}
\newcommand{\BOUniPHY}{Physics Department, Universit\`a degli Studi di Bologna, Bologna 40126, Italy}
\newcommand{\CAUniCHE}{Department of Mechanical, Chemical, and Materials Engineering, Universit\`a degli Studi, Cagliari 09042, Italy}
\newcommand{\CAUniEEE}{Department of Electrical and Electronic Engineering, Universit\`a degli Studi, Cagliari 09023, Italy}
\newcommand{\CAUniPHY}{Physics Department, Universit\`a degli Studi di Cagliari, Cagliari 09042, Italy}
\newcommand{\CAINFN}{INFN Cagliari, Cagliari 09042, Italy}
\newcommand{\Carleton}{Department of Physics, Carleton University, Ottawa, ON K1S 5B6, Canada}
\newcommand{\Campinas}{Physics Institute, Universidade Estadual de Campinas, Campinas 13083, Brazil}
\newcommand{\CentroFermi}{Museo della fisica e Centro studi e Ricerche Enrico Fermi, Roma 00184, Italy}
\newcommand{\CERNaddress}{CERN, European Organization for Nuclear Research 1211 Geneve 23, Switzerland, CERN}
\newcommand{\CIEMAT}{CIEMAT, Centro de Investigaciones Energ\'eticas, Medioambientales y Tecnol\'ogicas, Madrid 28040, Spain}
\newcommand{\Cluj}{National Institute for R\&D of Isotopic and Molecular Technologies, Cluj-Napoca, 400293, Romania}
\newcommand{\CPPM}{Centre de Physique des Particules de Marseille, Aix Marseille Univ, CNRS/IN2P3, CPPM, Marseille, France}

\newcommand{\CTLNS}{INFN Laboratori Nazionali del Sud, Catania 95123, Italy}
\newcommand{\ENUniCEE}{Engineering and Architecture Faculty, Universit\`a di Enna Kore, Enna 94100, Italy}
\newcommand{\ETHZ}{Institute for Particle Physics, ETH Z\"urich, Z\"urich 8093, Switzerland}
\newcommand{\FNALaddress}{Fermi National Accelerator Laboratory, Batavia, IL 60510, USA}
\newcommand{\FortLewis}{Department of Physics and Engineering, Fort Lewis College, Durango, CO 81301, USA}
\newcommand{\GEUni}{Physics Department, Universit\`a degli Studi di Genova, Genova 16146, Italy}
\newcommand{\GEINFN}{INFN Genova, Genova 16146, Italy}

\newcommand{\Hawaii}{Department of Physics and Astronomy, University of Hawai'i, Honolulu, HI 96822, USA}
\newcommand{\Houston}{Department of Physics, University of Houston, Houston, TX 77204, USA}
\newcommand{\IHEPaddress}{Institute of High Energy Physics, Beijing 100049, China}
\newcommand{\IPNO}{Institut de Physique Nucl\`eaire d'Orsay, 91406, Orsay, France}
\newcommand{\INSTM}{Interuniversity Consortium for Science and Technology of Materials, Firenze 50121, Italy}

\newcommand{\JINR}{Joint Institute for Nuclear Research, Dubna 141980, Russia}
\newcommand{\Krakow}{M. Smoluchowski Institute of Physics, Jagiellonian University, 30-348 Krakow, Poland}
\newcommand{\Kurchatov}{National Research Centre Kurchatov Institute, Moscow 123182, Russia}

\newcommand{\Laurentian}{Department of Physics and Astronomy, Laurentian University, Sudbury, ON P3E 2C6, Canada}
\newcommand{\LNFINFN}{INFN Laboratori Nazionali di Frascati, Frascati 00044, Italy}

\newcommand{\Lodz}{Institute of Applied Radiation Chemistry, Lodz University of Technology, 93-590 Lodz, Poland}
\newcommand{\LPNHE}{LPNHE, CNRS/IN2P3, Sorbonne Universit\'e, Universit\'e Paris Diderot, Paris 75252, France}

\newcommand{\Manchester}{The University of Manchester, Manchester M13 9PL, United Kingdom}
\newcommand{\MEPhI}{National Research Nuclear University MEPhI, Moscow 115409, Russia}

\newcommand{\MIINFN}{INFN Milano, Milano 20133, Italy}
\newcommand{\MIPoliICA}{Civil and Environmental Engineering Department, Politecnico di Milano, Milano 20133, Italy}
\newcommand{\MIPoliCHE}{Chemistry, Materials and Chemical Engineering Department ``G.~Natta", Politecnico di Milano, Milano 20133, Italy}
\newcommand{\MIPoliEIB}{Electronics, Information, and Bioengineering Department, Politecnico di Milano, Milano 20133, Italy}
\newcommand{\MIPoliENE}{Energy Department, Politecnico di Milano, Milano 20133, Italy}
\newcommand{\MIUni}{Physics Department, Universit\`a degli Studi di Milano, Milano 20133, Italy}
\newcommand{\MSU}{Skobeltsyn Institute of Nuclear Physics, Lomonosov Moscow State University, Moscow 119234, Russia}
\newcommand{\NAINFN}{INFN Napoli, Napoli 80126, Italy}
\newcommand{\NAUniPHY}{Physics Department, Universit\`a degli Studi ``Federico II'' di Napoli, Napoli 80126, Italy}
\newcommand{\NAUniCHE}{Chemical, Materials, and Industrial Production Engineering Department, Universit\`a degli Studi ``Federico II'' di Napoli, Napoli 80126, Italy}
\newcommand{\NAUniPHARM}{Pharmacy Department, Universit\`a degli Studi ``Federico II'' di Napoli, Napoli 80131, Italy}
\newcommand{\NAUniEEIT}{Department of Electrical Engineering and Information Technology, Universit\`a degli Studi ``Federico II'' di Napoli, Napoli 80125, Italy}
\newcommand{\NSU}{Novosibirsk State University, Novosibirsk 630090, Russia}

\newcommand{\Petersburg}{Saint Petersburg Nuclear Physics Institute, Gatchina 188350, Russia}
\newcommand{\PGUniCBB}{Chemistry, Biology and Biotechnology Department, Universit\`a degli Studi di Perugia, Perugia 06123, Italy}
\newcommand{\PGINFN}{INFN Perugia, Perugia 06123, Italy}
\newcommand{\PIINFN}{INFN Pisa, Pisa 56127, Italy}
\newcommand{\PIUniPHY}{Physics Department, Universit\`a degli Studi di Pisa, Pisa 56127, Italy}
\newcommand{\PNNLaddress}{Pacific Northwest National Laboratory, Richland, WA 99352, USA}
\newcommand{\Princeton}{Physics Department, Princeton University, Princeton, NJ 08544, USA}
\newcommand{\Queens}{Department of Physics, Engineering Physics and Astronomy, Queen's University, Kingston, ON K7L 3N6, Canada}
\newcommand{\RHUL}{Department of Physics, Royal Holloway University of London, Egham TW20 0EX, UK}
\newcommand{\RMTreINFN}{INFN Roma Tre, Roma 00146, Italy}
\newcommand{\RMTreUni}{Mathematics and Physics Department, Universit\`a degli Studi Roma Tre, Roma 00146, Italy}
\newcommand{\RMUnoINFN}{INFN Sezione di Roma, Roma 00185, Italy}
\newcommand{\RMUnoUni}{Physics Department, Sapienza Universit\`a di Roma, Roma 00185, Italy}
\newcommand{\SAINFN}{INFN Salerno, Salerno 84084, Italy}
\newcommand{\SNOLABaddress}{SNOLAB, Lively, ON P3Y 1N2, Canada}
\newcommand{\SSUniCHP}{Chemistry and Pharmacy Department, Universit\`a degli Studi di Sassari, Sassari 07100, Italy}
\newcommand{\Sussex}{Physics and Astronomy, University of Sussex, Brighton BN1 9QH, UK}
\newcommand{\Temple}{Physics Department, Temple University, Philadelphia, PA 19122, USA}
\newcommand{\TNFBK}{Fondazione Bruno Kessler, Povo 38123, Italy}
\newcommand{\TNTIFPA}{Trento Institute for Fundamental Physics and Applications, Povo 38123, Italy}
\newcommand{\TNUni}{Physics Department, Universit\`a degli Studi di Trento, Povo 38123, Italy}
\newcommand{\TOINFN}{INFN Torino, Torino 10125, Italy}
\newcommand{\TOPoli}{Department of Electronics and Communications, Politecnico di Torino, Torino 10129, Italy}

\newcommand{\TRIUMFaddress}{TRIUMF, 4004 Wesbrook Mall, Vancouver, BC V6T 2A3, Canada}
\newcommand{\TUM}{Physik Department, Technische Universit\"at M\"unchen, Munich 80333, Germany}
\newcommand{\UB}{Universiatat de Barcelona, Barcelona E-08028, Catalonia, Spain} 
\newcommand{\UCDavis}{Department of Physics, University of California, Davis, CA 95616, USA}
\newcommand{\UCLA}{Physics and Astronomy Department, University of California, Los Angeles, CA 90095, USA}
\newcommand{\UMass}{Amherst Center for Fundamental Interactions and Physics Department, University of Massachusetts, Amherst, MA 01003, USA}
\newcommand{\UNAM}{Instituto de F\'isica, Universidad Nacional Aut\'onoma de M\'exico (UNAM), M\'exico 01000, Mexico}
\newcommand{\UOC}{Department of Chemistry, University of Crete, P.O. Box 2208, 71003 Heraklion, Crete, Greece}
\newcommand{\USP}{Instituto de F\'isica, Universidade de S\~ao Paulo, S\~ao Paulo 05508-090, Brazil}
\newcommand{\VTech}{Virginia Tech, Blacksburg, VA 24061, USA}
\newcommand{\Zaragoza}{Centro de Astropart\'iculas y F\'isica de Altas Energ\'ias, Universidad de Zaragoza, Zaragoza 50009, Spain}
\newcommand{\ARAID}{ARAID, Fundaci\'on Agencia Aragonesa para la Investigaci\'on y el Desarrollo, Gobierno de Arag\'on, Zaragoza 50018, Spain}

\begin{document}
\title{Design and construction of a  new detector to measure ultra-low radioactive-isotope contamination of argon}

\author{The DarkSide-20k Collaboration:}
\author[1]{C.~E.~Aalseth,}
\author[2]{S.~Abdelhakim,}
\author[3,4]{F.~Acerbi,}
\author[5]{P.~Agnes,}
\author[6]{R.~Ajaj,}
\author[7]{I.~F.~M.~Albuquerque,}
\author[1]{T.~Alexander,}
\author[8,9]{A.~Alici,}
\author[10]{A.~K.~Alton,}
% ...
\affiliation[1]{\PNNLaddress}
\affiliation[2]{\IPNO}
\affiliation[3]{\TNFBK}
\affiliation[4]{\TNTIFPA}
\affiliation[5]{\Houston}
\affiliation[6]{\Carleton}
\affiliation[7]{\USP}
\affiliation[8]{\BOUniPHY}
\affiliation[9]{\BOINFN}
\affiliation[10]{\Augustana}
% ...
\affiliation[20]{\CIEMAT}
% ...



\abstract{
Large liquid argon detectors offer one of the best avenues for the detection of galactic weakly interacting massive particles (WIMPs) via their scattering on atomic nuclei.
The liquid argon target allows exquisite discrimination between nuclear and electron recoil signals via pulse-shape discrimination of the scintillation signals.
Atmospheric argon (AAr), however, has a naturally occurring radioactive isotope, \ar{}, a $\beta$ emitter of cosmogenic origin. For large detectors, the atmospheric \ar{} activity poses pile-up concerns. 
The use of argon extracted from underground wells, deprived of \ar{}, is key to the physics potential of these experiments. 
The DarkSide-20k dark matter search experiment will operate a dual-phase time projection chamber with 50 tonnes of radio-pure underground argon (UAr), that was shown to be depleted of \ar{} with respect to AAr by a factor larger than 1400. 
Assessing the \ar{} content of the UAr during extraction 
is crucial for the success of DarkSide-20k, as well as for future experiments of the Global Argon Dark Matter Collaboration (GADMC). 
This will be carried out by the \dia{} experiment, a small chamber made with extremely radio-pure materials that will be placed at the centre of the ArDM detector, in the Canfranc Underground Laboratory (LSC) in Spain. 
The ArDM LAr volume acts as an active veto for background radioactivity, mostly $\gamma$-rays from the ArDM detector materials and the surrounding rock.
This article describes the \dia{} project, including the chamber design and construction, and reviews the background required to achieve the expected performance of the detector.
}

\maketitle

\section{Introduction}
\label{sect:context}

Dark matter has played a fundamental role in the evolution of the universe at a cosmological scale, having a measurable impact in its expansion rate, and significantly contributing to large-scale structure formation. Nowadays, dark matter accounts for around 27\% of the energy-matter content of the universe. In spite of its abundance, what makes up the dark matter is yet to be understood.
Many candidates have been hypothesized, ranging from the primordial black holes predicted by General Relativity, to Weakly Interacting Massive Particles, that are predicted by many theories beyond the Standard Model of particle physics. Direct observation of dark matter is a major objective in modern fundamental physics.

Experiments searching for WIMPs intend to detect them via the signals (ionization, scintillation, and heat) they might leave when they undergo elastic scattering off atomic nuclei. Noble elements, like xenon and argon, are ideal active targets, as they are bright scintillators.
Environmental radioactivity can easily overwhelm WIMP signals if not carefully eliminated or discriminated against. 
The detection properties of liquid argon (LAr) are particularly favourable for the rejection of radioactive backgrounds that produce electron recoils (ERs). Indeed, there is significant difference between the time distribution of the scintillation signals produced by these interactions compared to that of nuclear recoil (NR) events. The DEAP-3600 experiment, with 3200~kg of LAr, has demonstrated an exceptional pulse shape discrimination (PSD) against such background, projected to be over $10^9$~\cite{Amaudruz:2017ekt}. The DarkSide-50 experiment has demonstrated the background-free capability of the dual-phase time projection chamber technique, in which both the primary scintillation and the electroluminescence from electrons multiplied in a gaseous region above the liquid are detected~\cite{Agnes:2018ep}. This technique also provides excellent position reconstruction capability that enables efficient fiducialization~\cite{Agnes:2015gu,Agnes:2016fz}. 
Single- or dual-phase LAr TPC with pulse-shape discrimination provides excellent sensitivity to WIMP masses above $\sim$30~\GeVcc{}. The dual-phase method also allows to search for lighter WIMP ($<$10~\GeVcc{}) using the electroluminescence signal alone, as demonstrated by DarkSide-50~\cite{Agnes:2018fg,Agnes:2018ft}. 
With careful control of ER background from local radioactivity and reduction of \ar{} background, a 1~tonne LAr detector has the potential to reach the \textit{neutrino floor} due to solar neutrino interactions, via coherent elastic scattering on target nuclei, in this low mass region.

Given the strong potential for the LAr technology to push the sensitivity for WIMP detection several orders of magnitude beyond current levels, scientists from 
ArDM, DarkSide-50, DEAP-3600, and MiniCLEAN have joined forces to found the Global Argon Dark Matter Collaboration (GADMC) to pursue a sequence of future experiments that will exploit this technology, starting with DarkSide-20k. 
A potentially limiting factor for the sensitivity of LAr-based experiments is the presence of the \ar{} radioactive isotope, present at a rate of 1~Bq/kg in atmospheric argon. 
One of the key enabling technologies of the GADMC program is the argon target obtained from the high-throughput extraction of low-radioactivity argon naturally depleted in \ar{} from underground sources (UAr) via the Urania plant in Cortez (USA). Urania will deliver 330~kg/day of 99.99\% purity UAr. The underground argon will be further chemically purified to detector-grade argon in Aria at the rate of 1~tonne/day. Aria is a 350~m cryogenic distillation plant currently being commissioned in Sardinia, Italy. 
Although not its primary goal, Aria can also be operated in isotope separation mode to achieve a 10-fold suppression of \ar{} per pass at a rate of 10~kg/day.

The \dia{} experiment at the Canfranc Underground Laboratory (LSC), Spain, will measure the \ar{} content in batches of the UAr delivered by Urania and Aria. In the shielded environment of the LSC, \dia{}  will be sensitive to very high depletion factors of \ar{}, of the order of 1000. These measurements are crucial for the DarkSide-20k physics program and for the future experiments of the Global Argon Dark Matter Collaboration.

\section{The \dia{} experiment}
\label{sect:intro}

\begin{figure}[!ht]
\centering
\includegraphics[height=0.55\textwidth]{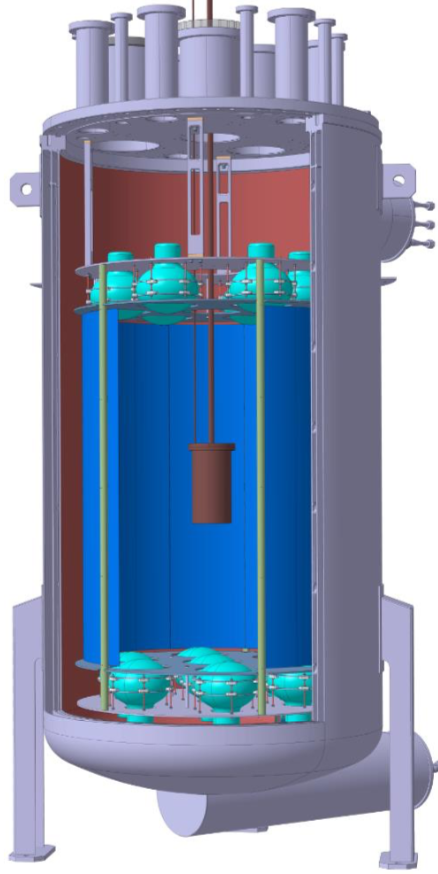}
\caption{DArT detector in the centre of the ArDM cryostat.}
\label{ardmdart}
\centering
\end{figure}

The \dia{} experiment at LSC aims at measuring UAr-to-AAr \ar{} depletion factors of the order of 1000 with 10\% precision in one week of running. The \ar{} isotope decays via $\beta$ emission with an end-point of 565~\keV{}. DArT is a single-phase liquid argon detector, with an active volume of approximately one litre, that will be filled with underground argon samples from Aria. Sampling directly UAr from Urania, with a SAES getter in front, would also be possible. The light produced by ionizing radiation in the active volume will be readout by two 1~cm$^2$ silicon photo-multipliers (SiPMs) procured  from the \DSk{} production packaged with the cryogenic readout electronics. DArT will be housed at the centre of the 1~tonne LAr ArDM detector~\cite{Marchionni:2011kg,Calvo:2016hve} (Fig.~\ref{ardmdart}), which will serve as an active veto to tag both internal and external radiation. The ArDM setup will be operated in the single-phase mode with a new set of low-radioactivity photo-multipliers (PMTs), 6 at the top and 7 at the bottom. A reflector foil coated with TPB wavelength shifter will cover the inner surface of the cylindrical barrel to enhance light collection. The cryostat is surrounded by a 50~cm thick polyethylene shielding (not shown in the Fig.~\ref{ardmdart}).

Background is largely composed of $\gamma$-rays originating from within the detector materials and from the surrounding cavern that are energetic enough to penetrate into the DArT chamber. To minimize the background induced by its intrinsic radioactivity, DArT is built with selected radio-pure materials. In DArT, the LAr is contained inside an outer vessel made of ultra-pure 99.99\% Oxygen Free High Conductivity (OFHC) copper. A smaller cylindrical structure made of radio-pure acrylic (PMMA) is inserted in the copper cell and provides the cleanest possible surface for the UAr under test, as well as the support for the SiPM detector assemblies. Following a procedure developed by the DEAP-3600 collaboration~\cite{Amaudruz:2012hr}, the inner surfaces of the acrylic structure directly in contact with the argon are sanded after fabrication to suppress backgrounds deriving from the plate-out of \rn\ daughters.

In order to suppress the impact of  external photons from the experimental cavern, which dominate the background budget, a 6~tonne lead shield will be installed around the ArDM vessel, in the hollow space between the ArDM cryostat and polyethylene shielding. The lead shield is a 140~cm height octagonal prism, with 10~cm width walls. In this configuration, the detector will be capable of measuring with high precision the larger \ar{} depletion factors of the UAr batches coming from the Urania plant.

\begin{figure}[!ht]
\centering
\includegraphics[width=0.41\textwidth]{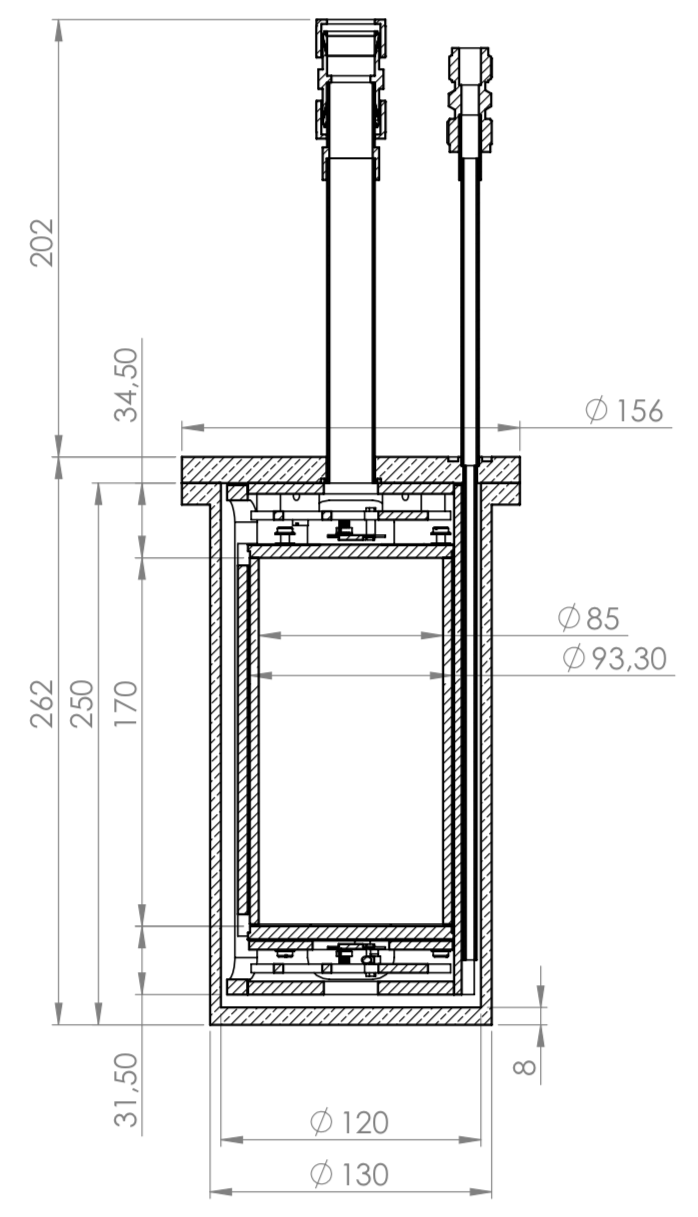}
\hspace{0.3 pt}
\includegraphics[width=0.42\textwidth]{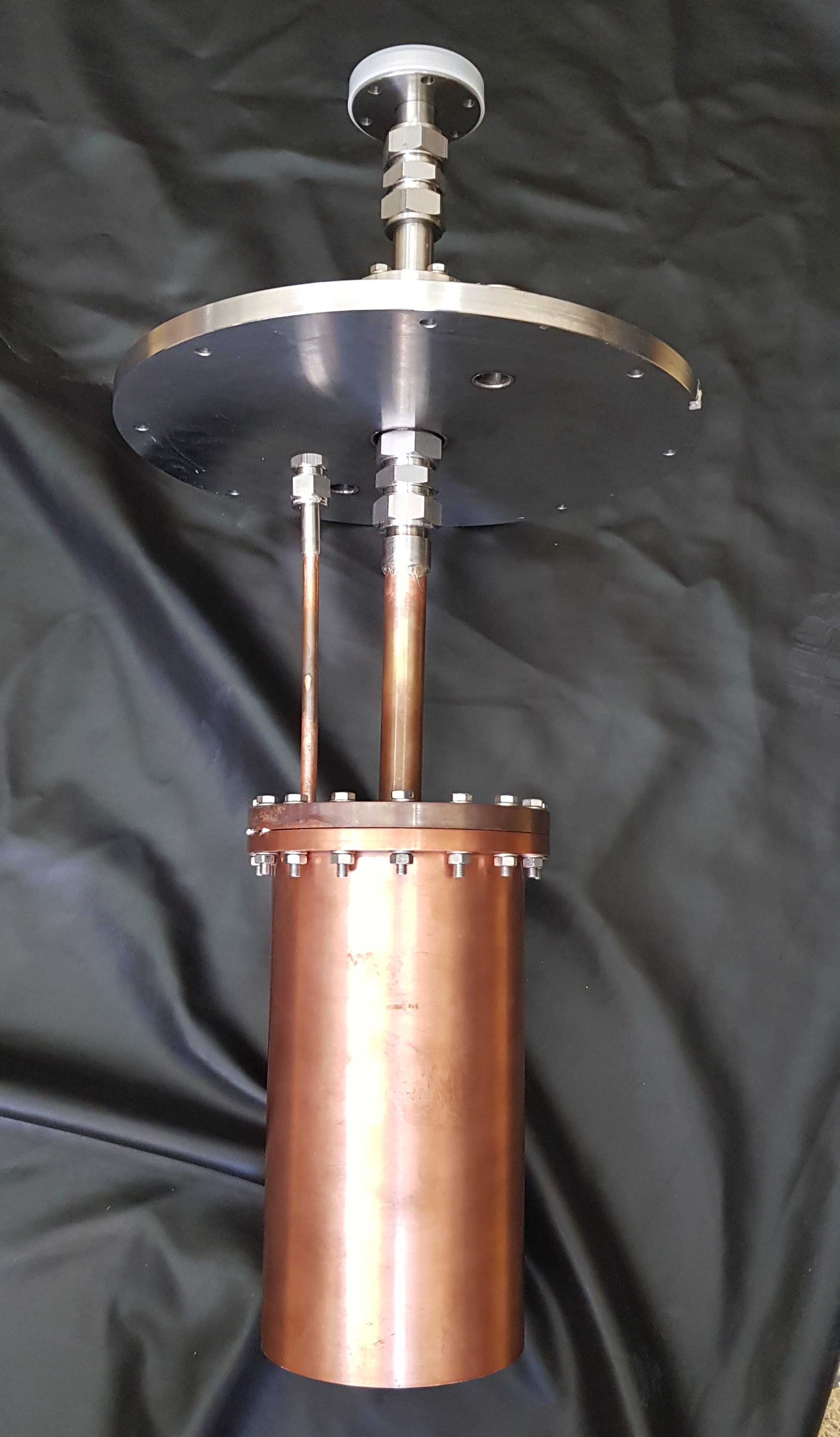}
\caption[]{(Left) Sketch of the inner structure of DArT. (Right) Picture of the actual detector. The short stainless steel pipes are used in the cryogenic tests, and will be replaced by longer tubes in the final setup.}
\label{mechanics:vessel}
\end{figure}

The lateral wall of the copper cylinder is 5~mm thick and the top and bottom caps are 8~mm thick (Fig.~\ref{mechanics:vessel}). The top flange allows access to the inner part of the chamber. The DArT instrumentation and services, including SiPM bias and readout, and a level meter signal, are routed through the top flange into a 15-cm long copper pipe leading to the top of the ArDM cryostat. A thinner off axis pipe, that penetrates the copper chamber and extends almost all the way to the bottom, is the argon fill line. 

\begin{figure}[!ht]
\centering
\includegraphics[width=0.35\textwidth]{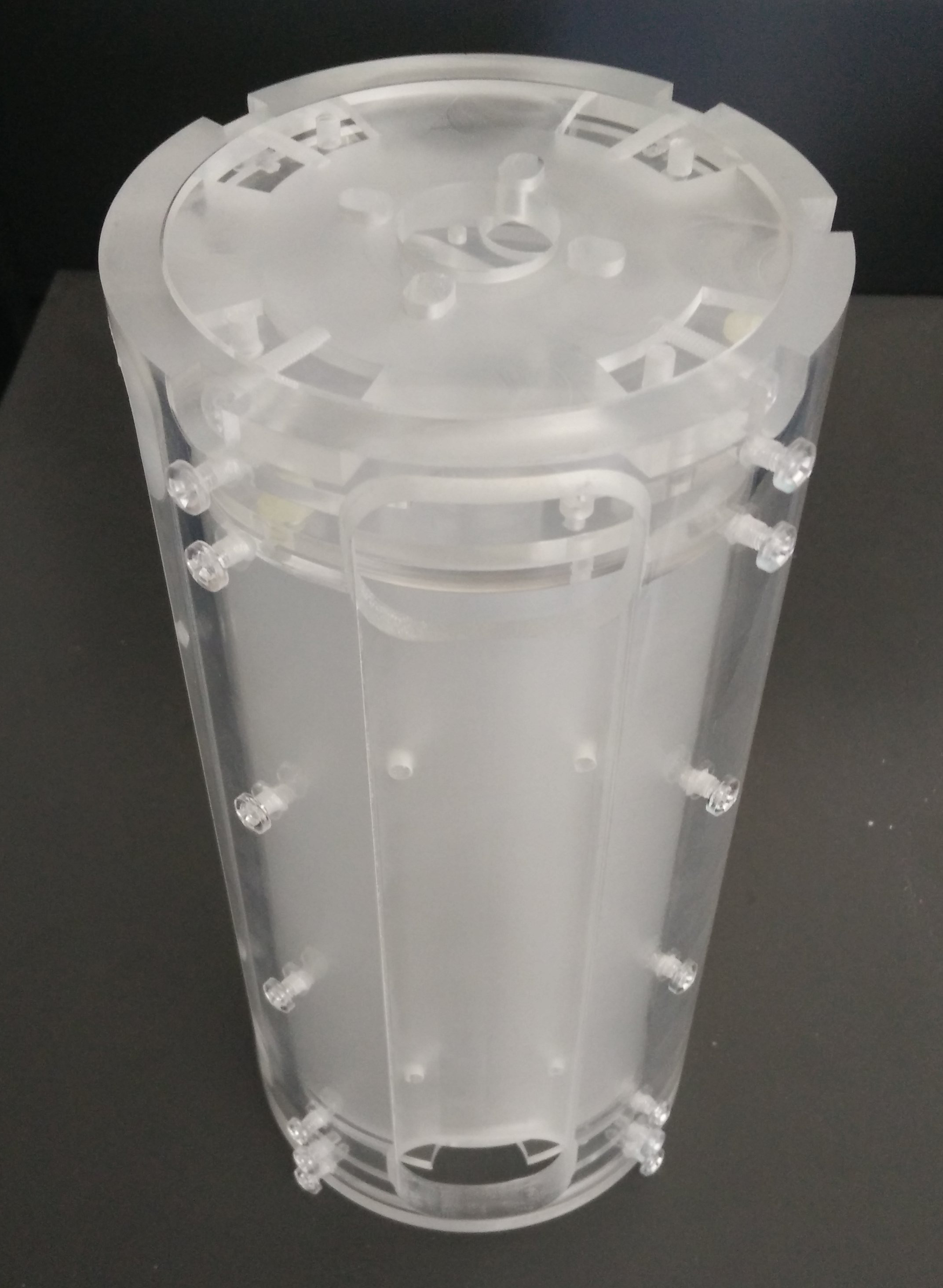}
\hspace{0.3 pt}
\includegraphics[width=0.5\textwidth]{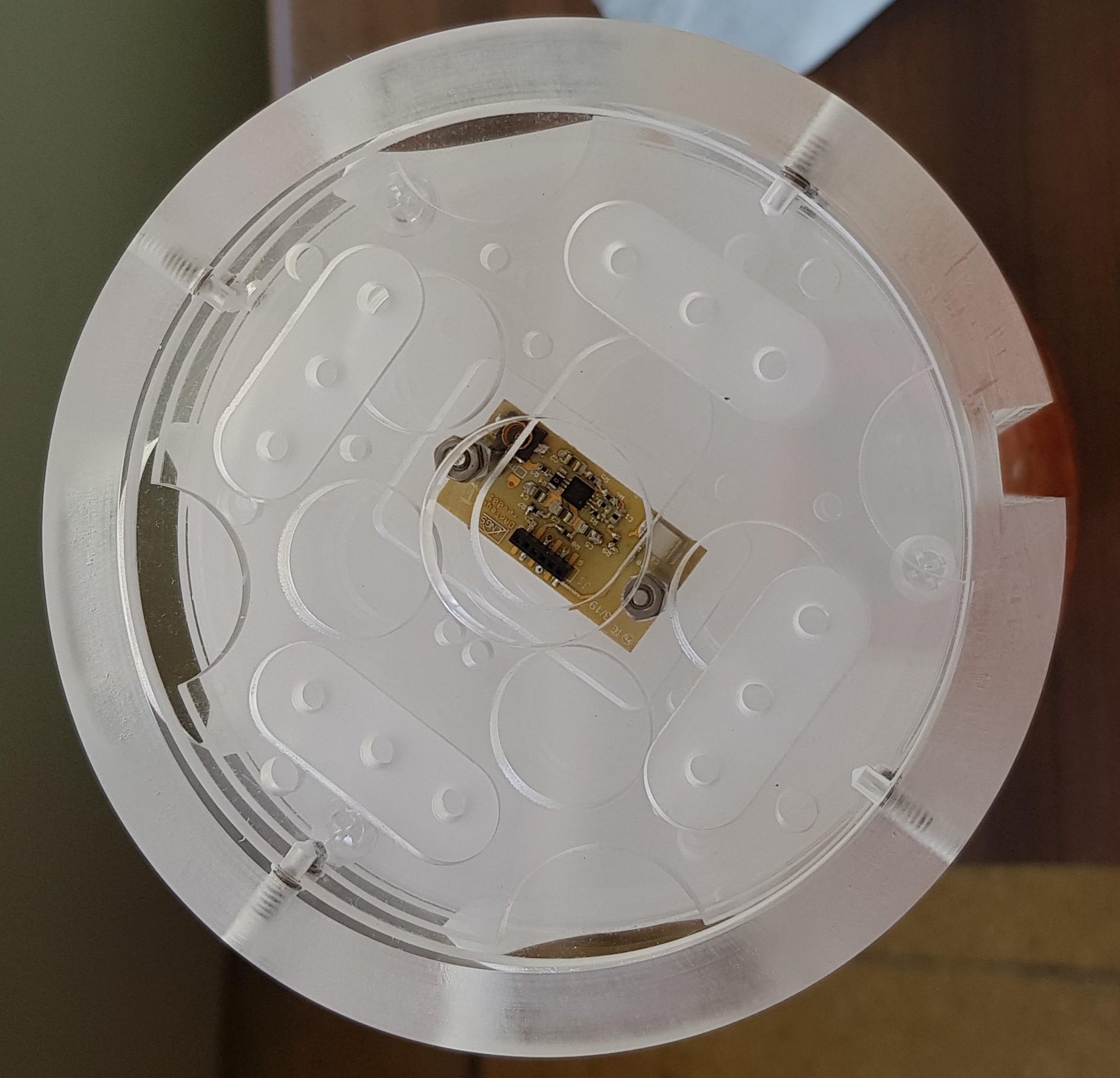}
\caption{(Left) Inner acrylic structure lateral view. (Right) Top acrylic cap and support disk with the photo-detector module installed.}
\label{photo:vessel}
\centering
\end{figure}

The inner acrylic structure is made of 8 pieces (Fig.~\ref{photo:vessel}): attachment to the top cover, inner and outer cylinders, 2 SiPM supports, 2 end caps, and facilities support on the bottom. Two 6~mm thick disks cap the inner cylinders in front of the SiPMs. The internal surfaces of these disks and the inner acrylic cylinder are coated with 200~$\mu$g/cm$^2$ wavelength shifter (TPB). The outer surface of the acrylic is wrapped in Vikuiti reflector foil~\cite{MInc:2010tq,Langenk_mper_2018}. An additional cylindrical shell (annulus) is used as a filler between the reflector foil and the copper vessel. The outer acrylic parts are attached to the top vessel flange. Two dedicated boards attached to the bottom and top acrylic disk with acrylic screws hold the SiPMs.

Currently in its construction phase, the \dia{} experiment is expected to test UAr not only for \DSk{}, but also for any future experiment of the GADMC, such as ARGO~\cite{Aalseth:2018gq} and a possible tonne-scale detector to search for light dark matter. 
It is also being considered for use by other experiments~\cite{Alexander:2019ux}.

\section{Cryogenics}

\begin{figure}[!ht]
\centering
\includegraphics[width=0.9\linewidth]{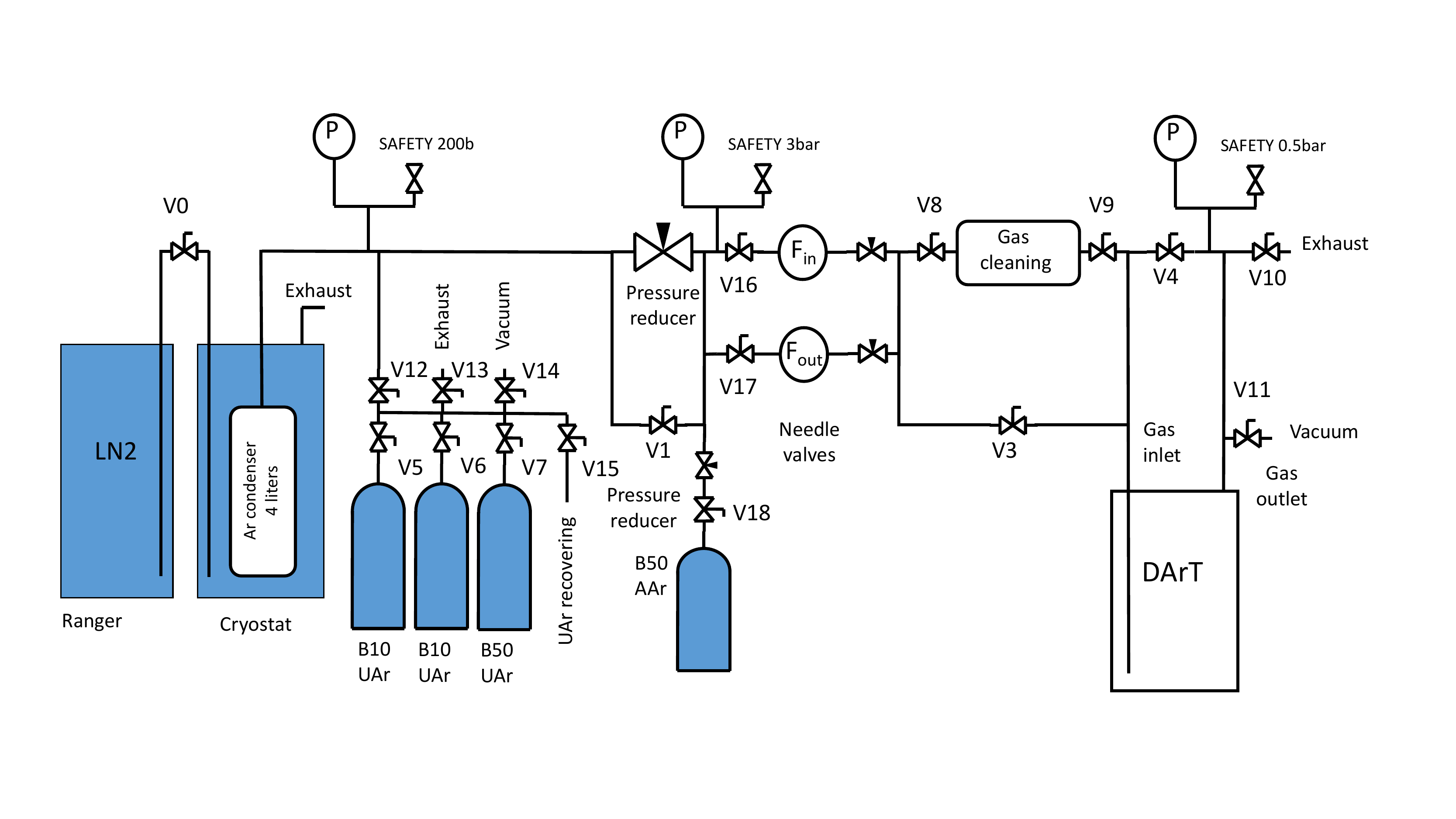}
\vspace*{-12mm}
\caption{Schematics of the gas handling and cryogenic system for the operation of DArT. P indicates mechanical pressure gauges, F stands for flow meters and V denotes valves. }
\label{dartoperation3}
\end{figure}

\dart{} is connected to the cryogenic system of the experiment through two copper pipes (Fig.~\ref{mechanics:vessel}). They are joined with a silver brazing to two stainless steel pipes that go up to a CF200 flange, on top of \ardm{}, which provides the external connections. The schematics of the gas handling system and cryogenics is in Fig.~\ref{dartoperation3}. High oxygen traces in the liquid argon would quench the scintillation signal produced in the detector target. Hence, less than 1~ppm of oxygen is left in the chamber before filling it with argon, pumping the system below 10$^{-3}$~mbar at room temperature. The pumping system is equipped with vacuum gauges and a turbo pump connected to valve V11. The pump is kept on while the system is warm. The system is proven leak tight at a level better than 10$^{-7}$~mbar~$\ell$/s.

In normal operational conditions the \dart{} chamber is kept isolated closing all the in/out valves and reaches the thermal equilibrium with the surrounding liquid argon of \ardm{}. The pressure inside \dart{} is very close to the pressure of \ardm{} which, in order to avoid argon contamination in case of a small leak, is kept at a few mbar above the atmospheric pressure.

\dart{} will undergo filling and emptying cycles as needed for testing different argon batches. The evacuated DArT chamber is filled by condensing gaseous argon on the walls of the vessel (5~mm thick) and of the copper pipes (1~mm thick), whose external sides are maintained at fixed temperature by the \ardm\ LAr bath. The filling rate is established by the inflow meter/controller, which supplies the STP argon gas to be cooled and condensed in DArT. Given the high thermal conductivity of copper, a substantial filling rate can be maintained without a significant drop of temperature across the copper walls and without a significant increment of the pressure inside \dart{}. The limiting factor for filling \dart{} is the ability of \ardm\ to dissipate safely the cooling and to condense heat to the cryocoolers. The \ardm\ dissipated heat is limited to a maximum of 20~W. The amount of LAr in \dart{} is 1.58~liters. To fill \dart{}, the energy needed for cooling the argon gas to its liquefaction point is 230~KJ, while 355~KJ are needed for its full condensation. The filling level of the DArT chamber is monitored by three PT1000 sensors. These platinum resistors are used as level sensors thanks to the different self-heating of their resistance in gas (with thermal conductance 3.2~mW/K) and liquid (32 mW/K) phases. When filling, the argon flows through a hot getter for its further purification before reaching the \dart{} chamber.

The time required for filling DArT is at least 8~hours. The expected run time of the experiment ranges from one to several weeks per argon batch, depending on the argon purity. Therefore, one full day for condensing and evaporating the argon in the chamber is not a dominant component in the duty cycle of the experiment.

To empty the \dart{} chamber, the argon is evaporated and temporarily stored in a 4~liter condenser bottle, enclosed in an LN$_2$ cryostat. As the condenser is refrigerated with liquid nitrogen (77~K), the argon converts into solid phase and therefore its saturated vapor pressure is about 260~mbar. This pressure cannot of course be applied directly to the liquid argon in the \dart{} chamber because the LAr would also freeze. The flow of gaseous argon into the condenser is therefore controlled in such a way that the pressure inside the \dart{} chamber is higher than the triple point pressure of LAr (682~mbar) and lower than the equilibrium pressure of \ardm{}. These operating conditions guarantee that neither the LAr will freeze within the \dart{} chamber nor the gaseous argon will re-condense on the walls on its way out. In particular, the LAr will become overheated and will evaporate on the free surface of the LAr, cooling a layer of liquid beneath the surface. Because of the low thermal conductivity of the LAr, the surface is thermally insulated from the bulk of LAr and the evaporation process is very inefficient. To speed up the evaporation of the LAr, two PT100 platinum resistors, dissipating 3.5~W each, are located near the bottom of the \dart{} chamber. The work required to evaporate the DArT LAr is 355~kJ, so the chamber is emptied in about 14~hours. When the \dart{} chamber is empty, the condenser will be full and it may be warmed up. The argon gas is then transferred to a high-pressure storage bottle.

\section{Photo-electronics}

Scintillation light in DArT will be collected by two photo-detector elements (PDE), specifically designed to maximise the radiopurity and simplify the connections. Each PDE is based on a module integrating both, a $11.9\times 7.8$~mm$^2$ SiPM, and the readout electronics in a $15\times 26$~mm$^2$ PCB. The SiPM cell size is 25~$\mu$m, and has quenching resistor of 10~M$\Omega$ at 77~K. These SiPMs are based on the NUV-HD-Cryo technology developed by FBK for DarkSide-20k, which allows low afterpulse at cryogenic temperature with extended over-voltage~\cite{Acerbi:2017gy}. The SiPM sensitivity peaks at 400-420~nm, with a photo-detection efficiency above 50\% at room temperature. The pre-amplifier is based on the low-noise design developed at LNGS~\cite{DIncecco:2018fx}.
Overall, a signal-to-noise ratio in excess of 40~is achieved when operated at an overvoltage of 7~V, with a rise time of 2~ns and a total power dissipation of 200~mW.
The dynamic range is measured in laboratory tests to be between 1.5 and 1000 photo-electrons~(PE). An optical fiber routed inside the LAr chamber will carry LED pulses to be used to monitor the response and the stability of the SiPMs over time.

The energy scale of DArT will be studied with AAr, that provides a well-known, high statistics \ar{} spectrum.  The energy resolution will be measured using an $^{83m}$Kr gaseous source injected into the argon stream, a method first demonstrated in argon in 2009~\cite{Lippincott:2010jb} and successfully exploited since then in many experiments, including DarkSide-50~\cite{Agnes:2018ep,Agnes:2017cz}.

\begin{figure}[!ht]
\centering
\includegraphics[width=0.60\textwidth]{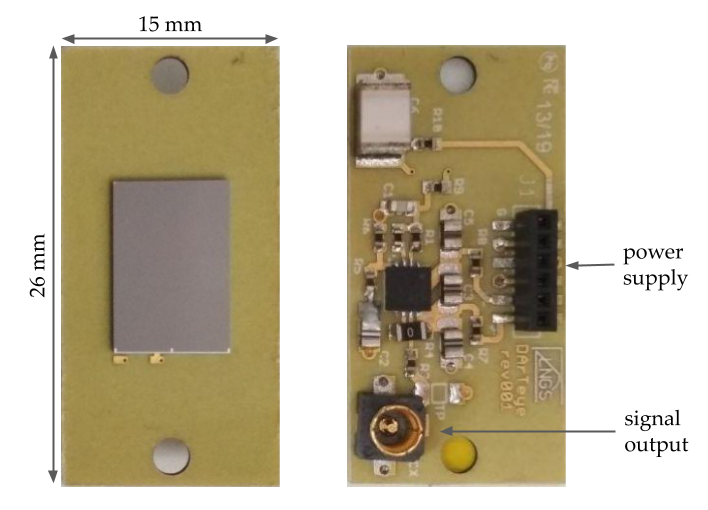}
\caption{Images of the front and back sides of the front-end boards housing the SiPM for light readout.}
\label{board_size}
\centering
\end{figure}

\section{Signal and background studies}
\label{sect:geometry}

In the DArT detector, signal events are electron recoils from the $\beta$ decay of \ar{}, which have an energy end-point of 565~\keV{} occurring within the inner acrylic vessel. These events deposit all their energy in DArT, leaving no signal in the veto detector, ArDM. The range of the energy spectrum in DArT below 600~\keV{} is defined as the \textit{region of interest} (ROI) where the signal events are contained. Background events stem from radioactive decays in the detector materials and in the experimental hall surrounding the detector. They typically produce $\gamma$ particles that deposit energy in DArT and/or in ArDM via Compton scattering. We assume that nuclear recoils are efficiently rejected using the powerful scintillation pulse shape discrimination technique in LAr. Background events that leave a signal with energy in the DArT ROI are tagged as background and removed from the analysis when they also deposit more than 10~\keV{} in ArDM. 
If, however, less than 10~\keV{} is deposited in ArDM they will go \textit{untagged} and contribute to the background of the measurement.

The amount of \ar{} in an argon sample is determined from fits to the spectral shapes of electron-like recoils in the LAr target with no coincident ArDM tag. The normalization of the background can be  constrained by the high-energy region of the DArT energy spectrum, above 600~\keV{}, where no signal events reside and whose rate is assumed constant or predictable in time.

Studies of the physics reach of the experiment are performed with tuned-on-data Monte Carlo simulations based on Geant4~\cite{Allison:2006cd}, with the simulation framework of the \DSf{} experiment (G4DS~\cite{Agnes:2017cz}). This package contains a detailed description of DArT and ArDM layout, including the lead and polyethylene shields. Predictions for the number of signal and background events expected in the ROI are based on these simulations. Background rates and origin are based on actual assays of screened materials and {\it in situ} measurements in the underground hall A at LSC. The values for the radio-impurities and masses of internal materials are summarized in Table~\ref{table1}, along with the figures for the external background (from outside ArDM).

\begin{figure}[!ht]
\includegraphics[width=0.55\textwidth]{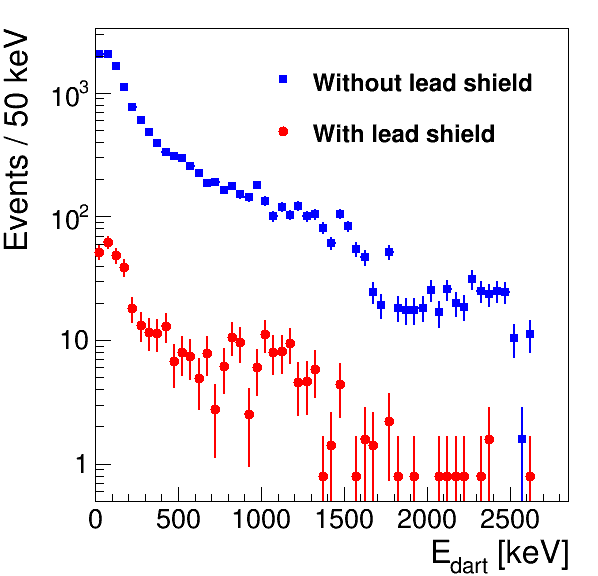}
 \centering
\caption{Expected untagged background events in \dart{} with the lead shield (red circles) and without (blue squares) in one month of data taking, assuming a veto threshold of 10~keV.}
\label{total_bakcground_dart}
\end{figure}

A study was conducted to assess the need for the lead shield. The total number of background events per week in the ROI with the lead shield is 7210, of which 465 are untagged. About 35\% of the background comes from sources external to ArDM~\cite{Calvo:2017qre}. For comparison, the expected number of untagged background events in the ROI without the lead shield is 10300 out of 122000 total events (96\% external background, see Fig.~\ref{total_bakcground_dart}).

The DArT fiducial mass is 1.42~kg of LAr. Assuming the \ar{} activity in the UAr measured by DarkSide-50 is 0.73~mBq/kg (for a corresponding depletion factor of 1400), we expect 631~signal events per week, yielding a signal-to-background ratio (S/B) of $\approx 0.06$ without the lead shield. The lead shield reduces the external background by a factor $\sim$20, leading to S/B $\approx 1.3$.

\begin{table}[!ht]
\centering
\caption[Radio-impurities of materials and external fluxes]{Activities (in mBq/kg, except for $^{210}$Bi that is in mBq/m$^2$), masses (kg) and surface areas (cm$^2$) for the different materials used for ArDM and DArT simulations. As the $\rm ^{238}$U secular equilibrium in Arlon is broken, the upper, middle and lower parts of the chain are considered. (*) $\gamma$ flux per~cm$^2$ at the outer surface of the polyethylene shielding of ArDM. (**) Surface in~cm$^2$.\vspace{3mm}}
\begin{tabular}[c]{|c|c|c|c|c|c|c|c|c|}  \hline
 \multicolumn{2}{|c|}{ \textbf{Source}} &  \textbf{   $\rm ^{238}$U   }  &  \textbf{   $\rm ^{232}$Th   }  &  \textbf{   $\rm ^{40}$K   }  &  \textbf{   $\rm ^{60}$Co   }  &  \textbf{   $\rm ^{210}$Pb   }  &  \textbf{   $\rm ^{210}$Bi   }  & \textbf{  mass [kg]  } \\
 \hline\hline
 \multicolumn{2}{|c|}{ \ardm{} Cryo} & 3.42 & 6.37 & 1.3 & 11.21 &  & &1630\\ \hline
      &  Base & 9176 & 11043 & 1978 &  &  &  &\\
      ArDM  &  Metal & 181 & 73 & 371 &  &  & & 10.8\\
      PMTs   &  Glass & 636 & 115 & 53 & 2 &  & &\\ \hline
\multicolumn{2}{|c|}{ Lead shield} & 0.37 & 0.073 & 0.31  &  & 10 & & 6000\\  \hline
\multicolumn{2}{|c|}{ \ardm{} pillars}&   0.012 & 0.04 & 0.06  & 0.04 & & & 26.5\\  \hline
\multicolumn{2}{|c|}{ \ardm{} rings}&   3.42 & 6.37 & 1.3 & 11.21 &  &   & 4\\  \hline
 \multicolumn{2}{|c|}{ PMTs support}  & 3.42 & 6.37 & 1.3 & 11.21 &  & & 16\\   \hline
 \multicolumn{2}{|c|}{ Acrylic} & 0.004 & 0.005 &   &  &  & 0.22 & 1.8\\  \hline
\multicolumn{2}{|c|}{ \dart{} Cu} &   0.012 & 0.04 & 0.06  & 0.04 & & & 6.95\\  \hline 
    &   Up  & 3.8 &  & & & & &\\
    SiPM &  Mid & 53& 70 & 1300& & & & 0.001\\
    Arlon & Low & 137 &  &  & & & &\\
 \hline
 \multicolumn{2}{|c|}{ Solder brazing} & 1203 & 406 & 3090 & & & & 0.001\\
 \hline
 \hline

\multicolumn{2}{|c|}{ External } &   0.72* & 0.13* & 0.05*  &  & & & 800700**\\
 \hline

\end{tabular}
\label{table1}
\end{table}

The current \ardm\ single-phase geometry, with only 13 PMTs and \dart{} shading some of the light, will be able to detect events with energies as low as 10~\keV{}. The sensitivity to the \ar{} signal is subject to further optimisation by tuning the ArDM veto energy threshold. The dependence of the number of untagged background events in DArT is calculated as function of such threshold, and is found to be linear with a slope of $8.5\times 10^{-3}/\keV{}$.

\section{Light response simulation}

The energy deposits in the liquid argon of \dart{} produce vacuum ultraviolet photons (VUV) that, once converted to visible photons (420~nm) by the TPB, are either detected by the two SiPMs or absorbed elsewhere. The simulation of the propagation of photons to the SiPMs with Geant4 takes into account the optical properties of the detector materials and their interfaces, in particular the acrylic (PMMA), the TPB coating of the internal surfaces, the reflector foil and the SiPM planes. Most of these materials are modelled as pure dielectrics, with the exception of the reflector which is assumed metallic, {\it i.e.} a surface that photons cannot penetrate. The TPB is assumed to re-emit a single visible photon (VIS) for each absorbed VUV photon with a characteristic time of 1.5~ns. The SiPMs are modelled as dielectrics with an arbitrarily reduced absorption length, in order to fully absorb the transmitted visible photons in a few nanometers. Absorbed photons are converted to photo-electrons, with a photon detection efficiency (PDE) of 40\%. Simulation parameters are tuned according to reference~\cite{Agnes:2017cz}.

The light response of the detector is evaluated simulating 20000 \ar{} events in the DArT active volume. The mean energy of the \ar{} $\beta$ emission is 210~\keV{}. Thus, considering a w-value for producing a VUV scintillation photon in LAr of 19.5\eV{}, and a quenching factor of 0.95 for electrons~\cite{Mei:2008ca}, on average 10230 photons are produced per event, corresponding to 48.7 VUV-photons\keV{}. The fraction of visible photons produced per VUV photon is $\sim$98.5\%, which accounts for the tiny absorption of VUV photons in LAr.

\begin{figure}[!ht]
\includegraphics[width=0.49\textwidth]{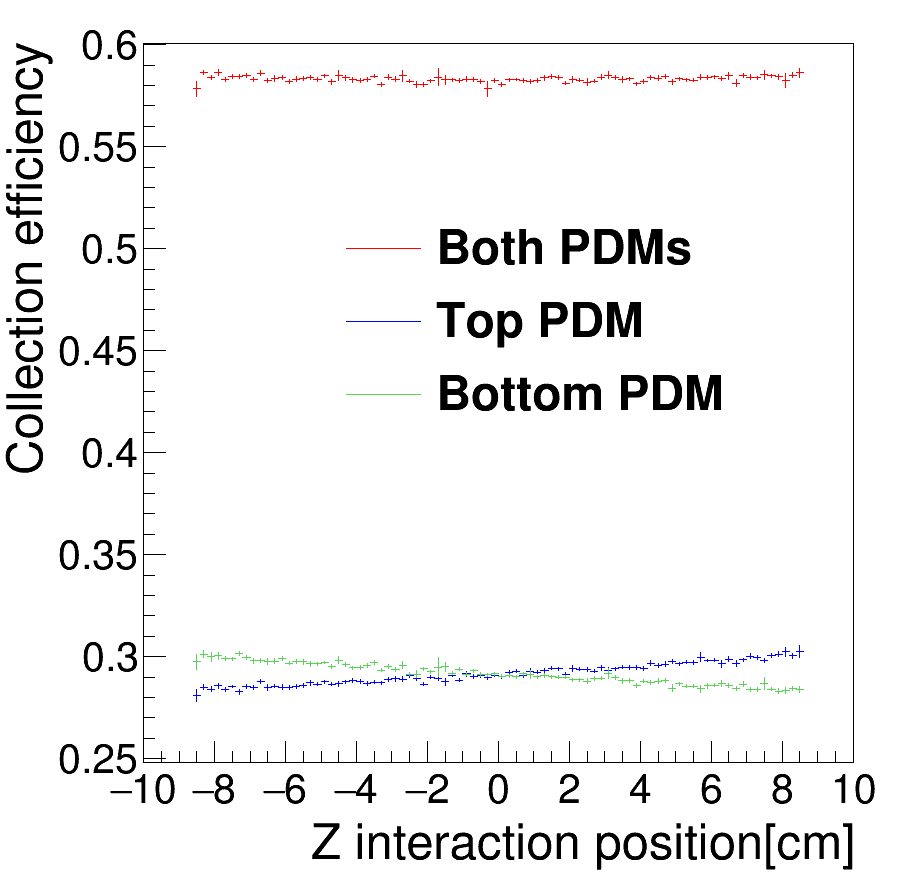}
\includegraphics[width=0.49\textwidth]{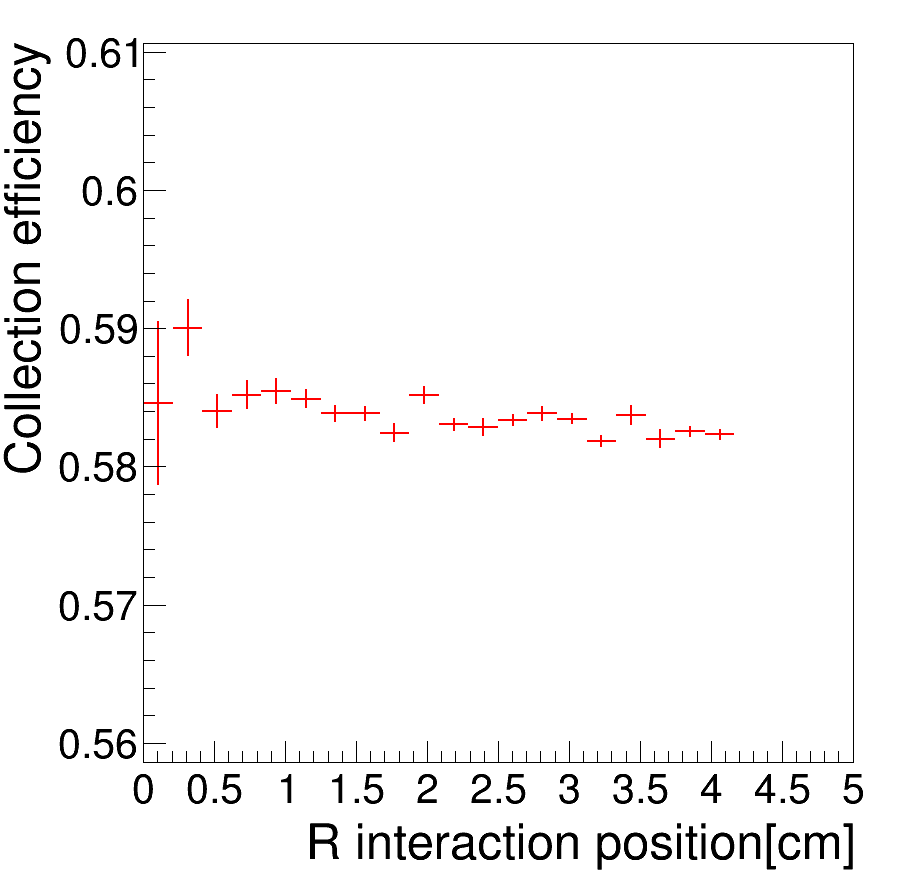}
 \centering
\caption[Collection uniformity in \dart]{Average light collection efficiency vs. height (left) and radius (right) of the interaction position.} 
\label{uniformity}
\end{figure}

The reflector foil is placed on the sides of the active volume and on the top and bottom acrylic caps, around the SiPMs, providing a light collection efficiency largely independent of the event position (Fig.~\ref{uniformity}). The average light collection efficiency is $\sim$58\%, which corresponds to a light yield of 11 PE/\keV{} for 40\% PDE. For reference, about 3000 PE are recorded per SiPM for energy deposits at the endpoint of the \ar{} spectrum.
Approximately, 30\% of the scintillation photons are collected in $\sim$4~ns (the fast component in Ar), that is $\sim$900 PE, well within the dynamic range of the readout electronics. The slow scintillation component delivers photons over an extended period. If the dynamic range of the light sensors turns out to be insufficient, the LY can be decreased by reducing either the overvoltage of the SiPMs (that is related to the PDE) or the effective area of the reflector.

The energy resolution is calculated with energy deposits from 50~keV up to 800~keV uniformly distributed in the \dart{} active volume. An energy resolution between 3 \%  and  6 \%  is found in the ROI.

\section{Sensitivity to the \boldmath \ar{} signal}

Using the background values and the light characterization illustrated above, we evaluate the \dart{} sensitivity expected for different \ar{} depletion factors, using a veto threshold of 10~\keV{}. In the following, we calculate both the measurement uncertainty on the depletion factors in case a clear signal is observed and the 90\% confidence level (C.L.) upper limits in absence of an excess over the expected background.

\begin{figure}[!ht]
\includegraphics[width=0.49\textwidth]{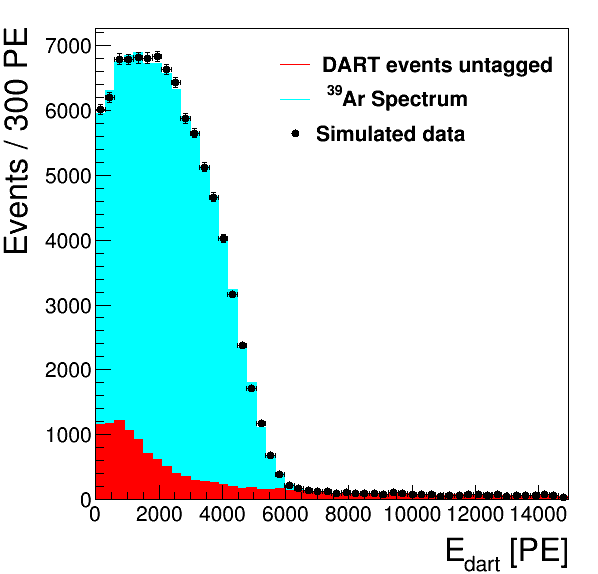}
\includegraphics[width=0.49\textwidth]{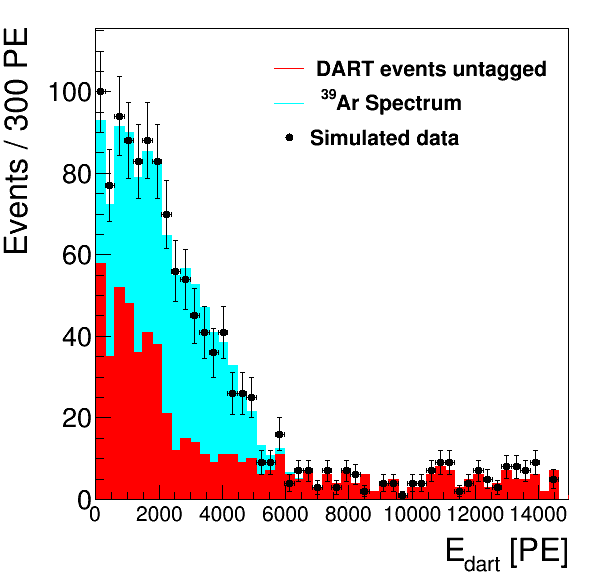}
\centering
\caption{Photo-electron spectra corresponding to one week of data taking, (left) for an \ar{} $DF = 10$ without lead shield and (right) for $DF = 1400$ with lead shield. The red (dark) histogram represents the background spectrum, the blue (light) histogram is the \ar\ signal and the black dots denote the simulated data.}
\label{fit_sensitivity}
\end{figure}

A typical photo-electron spectrum, for background and signal events acquired in one week of running of DArT, is displayed in Figs.~\ref{fit_sensitivity} for \ar{} depletion factors of 10 and 1400, respectively. The so-called \textit{simulated data} distribution is a randomized distribution generated from the sum of the signal and background distributions. The \ar{} content is extracted from a fit of the observed shape to the weighted sum of the signal and background distributions. From these fits, the expected statistical uncertainties per week on the measurement of different depletion factors of \ar{} are extracted. In the configuration with the lead shield, the uncertainty is below 1\% for a depletion factor ($DF$) of 10, 1\% for $DF = 100$, 7\% for $DF = 1400$ and 40\% for $DF = 14000$. The upper limit at the 90\% C.L. is reached for $DF \approx 6\times 10^4$. Without the lead shield, the statistical uncertainties increase, typically, by a factor 3, and the 90\% C.L. upper limit is $DF \approx 5000$.

These results assume the knowledge of the spectral shape of the signal and background from the simulations and, therefore, do not include systematic uncertainties. Systematic uncertainties on the \ar{} decay spectrum can be significantly reduced validating the simulated spectra with a large data sample in an \AAr{} run. 

The background shape in the ROI will also need to be validated with data. The analysis strategy envisaged is to start from the total measured background spectrum (before using the tagging cut) and then applying to it the veto cut efficiency vs. energy obtained from the simulations. From our relatively high threshold and detailed simulation studies, we are confident that the efficiency is high.

\section{Conclusions}

We have designed the \dia{} experiment capable of measuring with high precision the amount of \ar{} in the argon that will be used in current and future dark matter search experiments. The detector is under construction. The sensitivity of \dia{} was studied using a detailed background model and light response simulations. These studies show that our detector will allow us to measure UAr depletion factors in excess of 1000 with statistical accuracy better than 10\% 
in one week of counting time. Dedicated studies of systematic uncertainties on the background shapes are underway. 

\section*{Acknowledgments}

The DarkSide Collaboration would like to thank LNGS and its staff for invaluable technical and logistical support. This report is based upon work supported by the U. S. National Science Foundation (NSF) (Grants No. PHY-0919363, No. PHY-1004054, No. PHY-1004072, No. PHY-1242585, No. PHY-1314483, No. PHY- 1314507, associated collaborative grants, No. PHY-1211308, No. PHY-1314501, No. PHY-1455351 and No. PHY-1606912, as well as Major Research Instrumentation Grant No. MRI-1429544), the Italian Istituto Nazionale di Fisica Nucleare (Grants from Italian Ministero dell'Istruzione, Universit\`a, e Ricerca Progetto Premiale 2013 and Commissione Scientific Nazionale II). We acknowledge the financial support from the UnivEarthS Labex program of Sorbonne Paris Cit\'e (Grants ANR-10-LABX-0023 and ANR-11-IDEX-0005-02), the S\~ao Paulo Research Foundation (Grant FAPESP-2016/09084-0), and the Russian Science Foundation Grant No. 16-12-10369. The authors were also supported by the ``Unidad de Excelencia Mar\'{\i}a de Maeztu: CIEMAT - F\'{\i}sica de part\'{\i}culas'' (Grant MDM2015-0509), the Polish National Science Centre (Grant No. UMO-2014/15/B/ST2/02561), the Foundation for Polish Science (Grant No. TEAM/2016-2/17), the International Research Agenda Programme AstroCeNT (Grant No. MAB/2018/7) funded by the Foundation for Polish Science from the European Regional Development Fund, the Science and Technology Facilities Council, part of the United Kingdom Research and Innovation, and The Royal Society (United Kingdom). We also wish to acknowledge the support from Pacific Northwest National Laboratory, which is operated by Battelle for the U.S. Department of Energy under Contract No. DE-AC05-76RL01830.

\bibliographystyle{ds}
\bibliography{ds}

\end{document}